\documentclass[11pt]{article}
\usepackage[textwidth=16cm, textheight= 22cm]{geometry}

\usepackage{lineno,hyperref}

\usepackage{lineno}
\usepackage[numbers]{natbib}
\usepackage{graphics,multirow}
\usepackage{amsmath}
\usepackage{amssymb}
\usepackage{epsfig}
\usepackage{amsfonts}
\usepackage{color}
\usepackage{graphicx}
\usepackage[figuresright]{rotating}
\usepackage{array}
\usepackage{upgreek}
\usepackage{caption}

\renewcommand{\baselinestretch}{1.2}

\usepackage{lineno}

\def\beq{\begin{equation}}
\def\eeq{\end{equation}}

\begin{document}

\title{Integration of max-stable processes and Bayesian model averaging
to predict extreme climatic events in multi-model ensembles}

\author{ \bf{Yonggwan Shin}${}^{a}$,
  $~$\bf{Youngsaeng Lee}${}^{b}$,  $~$ \bf{Juntae Choi}${}^c$, \\
	and  $~$\bf{Jeong-Soo Park}${}^{d, *}$ 
	\\     
	\small\it a:  Data Analysis Group/Spatial Information Flatform Team, Neighbor System Inc., Korea. \\
	\small\it b: Digital Transformation Department, Korea Electric Power Corporation, Korea\\
	\small\it c: National Institute of Meteorological Science, Korea\\
	\small\it d: Department of Statistics, Chonnam National University, Gwangju 500-757, Korea\\
	\small\it *: Corresponding author, E-mail: jspark@jnu.ac.kr, Tel: +82-62-530-3445,\\ ORCID: 0000-0002-8460-4869
}

\maketitle 



\noindent{\bf ABSTRACT}\footnote[1]{{\it Stochastic Environmental Research and Risk Assessment}, 2019} \\

Projections of changes in extreme climate are {sometimes} predicted by using multi-model
ensemble methods such as Bayesian model averaging (BMA) embedded with the
generalized extreme value (GEV) distribution. BMA
is a popular method for combining the forecasts of individual simulation models by
weighted averaging and characterizing the uncertainty induced by simulating the
model structure. This method is referred to as the GEV-embedded BMA. It is,
however, based on a point-wise analysis of extreme events, which means it overlooks the spatial dependency between nearby {grid cells}.
Instead of a point-wise model, a spatial extreme model such as the max-stable process
(MSP) is often employed to improve precision by considering spatial
dependency. We propose an approach that integrates the MSP into BMA, which is
referred to as the MSP-BMA herein. The superiority of the proposed method over the
GEV-embedded BMA is demonstrated by using extreme rainfall intensity data on the Korean peninsula from Coupled Model Intercomparison Project Phase 5
(CMIP5) multi-models. The reanalysis data called APHRODITE (Asian Precipitation Highly-Resolved
Observational Data Integration Towards Evaluation, v1101) and 17 CMIP5 models are
examined for 10 grid boxes in Korea.
 In this example, the MSP-BMA achieves a variance reduction over the GEV-embedded BMA.
 The bias inflation by MSP-BMA over the GEV-embedded BMA is also discussed.
  A by-product technical advantage of the MSP-BMA
is that tedious `regridding' is not required before and after the analysis while it should be done for the GEV-embedded BMA. 

\section{Introduction}


 The multi-model ensemble has the potential to combine the strengths of individual models and to better represent forecast uncertainty than the use of a single model.
 Various studies have shown the advantage of merging forecasts from multiple climate models for climate projection
 (e.g., Casey 1995; Coelho et al. 2004; Stephenson et al. 2005; Kharin et al. 2013; Arisido et al. 2017).
A commonly used multi-model ensemble method is Bayesian model averaging
(BMA), which assumes a probability distribution around the forecasts of
individual models to establish the prior and uses a calibration period to determine
the static weights for each individual model (Parrish et al., 2012; Wang et al.,
2017).  

When our interest is in forecasting extreme climatic events, the generalized extreme
value distribution (GEVD) is typically used as an assumed probability
distribution in BMA. {The GEVD encompasses all three possible asymptotic extreme value distributions predicted by large sample theory (see, e.g., Leadbetter et al. 1983).}
 It is  widely used to analyze univariate extreme values (Coles, 2001).
 Zhu et al. (2013) performed a GEV-embedded BMA of extreme rainfall
intensities by using seven regional climate models (RCMs). Hereafter, this method is
referred to as the GEV-BMA. In that study, the method was applied to each grid cell separately; hence, it ignored the spatial dependency between
nearby grid cells. Consequently, the GEV-BMA may be less reliable for small samples and could
suffer from local fluctuations, which may be heavily influenced by outlier
observations. Thus, it may be better to generalize this separate GEVD approach to a
model taking into account the spatial dependency in the BMA framework. {Therefore} we consider a spatial extreme model, namely a
max-stable process {(see, e.g., de Haan 1984; Smith 1990; Davison et al. 2012)}. Max-stable process (MSP) is an infinite dimensional extension of the GEVD.
The approach considered here, an MSP-embedded BMA, is a spatial extension
of the GEV-BMA proposed by Zhu et al.(2013). Hereafter, the proposed method is referred to as the MSP-BMA.

MSPs are widely used for modeling spatial
extremes, which deal with the spatial dependency of extreme events arising
between nearby locations. {The MSP estimates not only the GEVD parameters as continuous functions of space but also spatial dependent structure.}
The advantage of the MSP is that it can be used to address problems concerning the aggregation of global processes over many locations.
{Because it builds a model for the whole set
of grid cells}, it avoids local fluctuations that may be heavily
influenced by outlier observations. Thus, more precise and robust return
levels of extreme events can be obtained.

The present study aims to determine whether the integration of the MSP and
BMA can add value when predicting future extreme values using multiple
simulation models. It was motivated by Parrish et al. (2012) and Madadgar and
Moradkhani (2014), who integrated data assimilation methods and copulas into
BMA, respectively. Their results demonstrated that the predictive distributions
after these integrations are more accurate and reliable, less biased, and more
confident (i.e., having lower uncertainty). Sabourin et al. (2013) considered using
BMA for multivariate extreme events. More recently, Najafi and Moradkhani
(2015) and Yan and Moradkhani (2016) considered spatial hierarchical Bayesian
modeling of extremes to obtain spatially distributed BMA weights corresponding
to each simulation model. {Our approach is in the same line of
integrating spatial models into BMA framework. We are unaware of any work in
which the MSP has been integrated into BMA. This is a reason why we
considered MSP-embedded-BMA approach  in this study.}
  Based on the success of their research, we also expect a
similarly good result by integrating the MSP and BMA.

 To demonstrate
the prediction ability of our approach, we use the annual maximum daily
precipitation
 as simulated by Coupled Model Intercomparison Project Phase 5 (CMIP5) models
  in historical and future experiments over the Korean peninsula.
Based on the variance estimates of return levels and other criteria, our approach seems outperforming the GEV-BMA.

The remainder of this paper is organized as follows. Section 2 describes models of extreme
events including the GEVD and the MSP. Section 3 explains the GEV-embedded
BMA and introduces the integrated MSP-BMA. Section 4
provides a numerical study of extreme rainfall
 as simulated by CMIP5 models over the Korean peninsula.
 A computationally faster version of the MSP-BMA is considered in Section 5.
 Section 6 concludes with discussion.

\section{Modeling extreme values}
\subsection{Extreme value distribution}
{The GEVD  is widely used to analyse univariate extreme
values. The three types of extreme value distributions are sub-classes of GEVD}. The cumulative distribution function of the GEVD
is as follows:
\begin{linenomath*}\begin{equation} \label{gevd}
G(x) = exp \left\{ - \left(1+ \xi {{x-\mu} \over {\sigma}}\right) ^{-1/\xi} \right\},
\end{equation}\end{linenomath*}
when $1+\xi (x- \mu ) / \sigma > 0$, where $\mu ,\; \sigma $, and $\xi$ are the
location, scale, and shape parameters, respectively. The particular case for
$\xi=0$ in  Eq. (\ref{gevd}) is the Gumbel distribution, whereas the cases for
$\xi>0$ and $\xi<0$ are known as the Fr\'{e}chet and the negative Weibull
distributions, respectively (\cite{coles}).

{Assuming the data, the annual maxima of daily precipitation in this study, follow  (approximately)  a GEV distribution}, the parameters can be
estimated by using the maximum likelihood method {(see, e.g., Wilks 1995 or Coles 2001)} or the method of L-moments
estimation (Hosking, 1990). The maximum likelihood estimator
is less efficient than the L-moments estimator in small samples for
typical shape parameter values (Hosking et al. 1985). The L-moments method
is used in this study because relatively short 30-year samples are analyzed for
each comparison period. Moreover, the formulae used to obtain the L-moments estimator are
simple compared with obtaining the maximum likelihood estimator, which needs an iterative computation until
convergence. Those are
\begin{equation}
\hat\xi = 7.8590c+2.9554c^2
\end{equation}
where
\begin{equation}
c=\frac{2 l_2}{l_3+3 l_2}-\frac{ln(2)}{ln(3)},
\end{equation}
and $l_2$ and $l_3$ are the sample second and third L-moments (Hosking,
1990). The other parameters are then given by
\begin{equation}
\sigma=\frac{\hat\xi \; l_2}{\Gamma(1+ \hat\xi )\;(1-2^{-\hat\xi })}
\end{equation}
 \begin{equation}
\mu=l_1+\frac{\hat\sigma [\Gamma(1+\hat\xi)-1]}{\hat\xi},
\end{equation}
where $l_1$ is the first sample L-moment.

{It can be helpful to describe changes in extremes in terms of changes in extreme quantiles.} These are obtained by inverting (\ref{gevd}):
\begin{equation}\label{z_p}
z_p = \mu - {\sigma \over \xi } [ 1- \{ - log (1-p) \} ^{-\xi} ],
\end{equation}
where $G( z_p )= 1-p $. Here, $z_p$ is known as the return level associated with
the return period $1/p$, since the level $z_p$ is expected to be exceeded on
average once every $1/p$ years (Coles, 2001). For example, a 20-year (50-year)
return level is computed as the 95th (98th) quantile of the fitted GEVD. This
return level is used as the main quantity in the numerical study to
compare the GEV-BMA with the MSP-BMA in the subsequent sections.

\subsection{Models for spatial extremes}
Extreme value distributions are typically applied independently for the datasets of each weather station (grid cell, in this study). However, a better approach may be to build a model for
all the weather stations together to avoid local fluctuations and reflect the
spatial dependency between observations from nearby stations.
 A variety of statistical tools have been used for the spatial modeling of extremes including Bayesian hierarchical models,
 copulas and max-stable processes. The first model is often referred to as ``latent variable" approach (Davision et al., 2012).

 Spatial hierarchical models introduce spatial variation in the parameters of the extremal response distributions.
 By letting the parameters of the response distributions depend on the unobserved latent processes, the spatial hierarchical model
 induces dependence in responses by integration over the latent process. A Gaussian process is frequently used for the latent process.
 This approach is common in geostatistics with nonnormal response variables (Diggle and Ribeiro, 2007), and because of the complexity
 of the integration involved is most naturally performed in a Bayesian setting, using Markov chain Monte Carlo algorithms to perform inferences
 (see, e.g., Sang and Gelfand, 2010; Cooley et al., 2012; Fawcett and Walshaw, 2016).  Compared to the models based on copulas and MSP,
  it is reported that the spatial hierarchical modeling makes quantifying uncertainty more straightforward, and it allows a better fit to marginal distribution,
  but it fits the joint distributions of extremes poorly (Davision et al., 2012).

Copula provides a unifying framework to modeling multivariate data, and so may be useful to model spatial extremes.
Given marginal GEVDs and some correlation functions, various copulas including Gaussian and Student t are available.
Some copulas satisfying the max-stability property, called as extremal copulas, may be useful to model spatial extremes
 (see, e.g., Nelsen 2006; Salvadori and De Michele 2004; Madadgar and Moradkhani 2014; Requena et al. 2016).

The max-stable processes (MSP) is an useful natural extension of the extremal (univariate and multivariate GEV) models.
 It is an infinite-dimensional generalization of the univariate GEV model, and is particularly applicable in the
context of a time series or a spatial process. This theory was introduced by De
Haan (1984) and has been extended by a number of researchers such as Smith
(1990), Schlather (2002), Kabluchko et al. (2009), and Padoan et al. (2010). Appropriately chosen max-stable models seem essential for successful spatial modeling of extremes (Davision et al., 2012).
It, however, also bring challenges. For example, the likelihood is hard to tractable for even moderate sample size problems, and consequently composite (or pairwise) likelihood methods are utilized.

For future projections and uncertainty assessment of extreme events from an ensemble of climate models, many proposed the integration of spatial extreme models and BMA framework. We are, however, unaware of any work in which the MSP has been integrated into BMA. Thus we propose an integration of MSP and BMA in multi-model ensemble, in this study.

\subsection{Max-stable processes}
\label{msp}
 The max-stable process (MSP) $Z(\cdot )$ is defined as the limit process of the maxima of the
independent and identically distributed random fields $Y_i (x), x \in  R^d$ (De
Haan, 1984) and can be expressed as
\begin{linenomath*}\begin{equation} \label{max-st}
Z(x) = \lim_{n \rightarrow  \infty} {{ \max_{i \le n} Y_i (x) -
b_n (x)} \over {a_n (x) }},
\end{equation}\end{linenomath*}
if a limit exists for all $x$ in $ R^d$ with normalizing
constants $a_n (x) > 0 $ and $b_n (x) \in R$.
For a fixed point ($x_1 , \cdots , x_d $) in space,
 any d-dimensional marginal
distribution of the MSP belongs to a class of multivariate extreme
value distributions. In practice, this means that the resulting
parameters $\mu (x),~\sigma(x) > 0$, and $\xi(x)$ are the
continuous functions to be estimated. Thus, the MSP naturally
permits modeling and predictions with data-level spatial
dependence. We then build the regression-based forms for
the parameters $ \mu(x),~\sigma(x),~\xi(x)$ and estimate the spatial
dependence parameters (\cite{westra}). These regression-based forms, however, may introduce some estimation bias for MSP while it are convenient to predict something for non-observed sites.

The precision of the
estimated parameters in the MSP is expected to be improved over the point-wise (or at-site) GEV model because the MSP allows more
data to be used in the model fitting by combining data from multiple sites across a
spatial domain (Ribatet, 2013; Gaume et al., 2013; and Oesting and Stein, 2017).

Given a series of $n$ observations at $K$ spatial locations, the aim of a
statistical analysis would be to fit an MSP by using assumed forms for the
parameters $\mu (x),~\sigma(x)$, and $\xi(x)$, while also estimating spatial
dependence. However, for more than $K=2$ spatial locations, the distribution
function of the general MSP has no analytically tractable form, which thereby
presents a problem for practical statistical model fitting. Therefore, ad hoc
characterization methods have been proposed. Smith (1990) introduced the first,
and Schlather (2002) and Kabluchko et al. (2009) later introduced other methods.
 Schlather's characterization involves a
correlation function $\rho(x)$, deriving a bivariate cumulative distribution function as
\begin{linenomath*}\begin{align}
V(z_1 , z_2)  = - \log \Pr \left\{ Z(x_1 ) \leq z_1 , Z(x_2) \leq
z_2 \right\} \cr
  =  {{1} \over{2}} \left( {{1} \over {z_1}}
+ {{1} \over {z_2}} \right) \times \left\{ 1+ \sqrt{ 1 - 2 ( \rho ( h) + 1 ) {{z_1 z_2 }
\over{ (z_1 + z_2) ^2 }}} \right\},\label{biv-dist}
\end{align}\end{linenomath*}
where $h$ is the distance between $x_1$ and $x_2$. Further, $\rho \in [0, 1]$,
where the lower bound corresponds to the independence of extremes. To take into account the spatial
dependency, we used a powered exponential correlation function  $\rho(h)$ of
distance $h$, in this study \cite{cressie11}:
\begin{linenomath*}\begin{equation}
\label{corrfun} \rho(h) =  \exp \left\{ - \left( {{h} \over {\tau}} \right) ^\eta \right\}, \;~
0 < \eta \le 2 , \;\tau > 0, \; h>0,
\end{equation}\end{linenomath*}
where $\tau$ and $\eta$ are the scale and shape parameters that control the
range and smoothness of the process, respectively. The Brown--Resnick model
\cite{Kabluchko} uses the following variogram  \cite{cressie11}:
\begin{equation}
\alpha^2 (h) = 2 (h / \tau )^\eta ,  \;~ 0 < \eta \le 2 , \;\tau > 0.
\end{equation}

The full likelihood of the MSP is often analytically unknown or computationally
prohibitive to calculate. The following pair-wise log likelihood function is used
because only the bivariate density function is specified;
\begin{linenomath*}\begin{equation} \label{CL}
 {l_p (\psi) = {\sum_{k=1} ^{n}}{\sum_{i <
\cdots < j }} \;log \; f_X ( x_{{i}(k)}, \cdots , x_{{j}(k)} ; \psi),}
\end{equation}\end{linenomath*}
where $x_{{i}(k)}$ is the $k$-th observation at the $i$-th site for $i \in \left\{ 1,
\cdots , d \right\}$.
 The use of the pair-wise likelihood method leads to an underspecified model.
Thus, instead of the Akaike information criterion, the Takeuchi information
criterion (TIC) is used to select a better model for the estimated parameters
$\hat\psi = (\hat\mu(x),~\hat\sigma(x),~\hat\xi(x))$  \cite{VarinVidoni}:
\begin{equation}
TIC= -2l_p ( \hat \psi ; y ) + 2\; tr \left( J (\hat\psi ) H (\hat\psi )^{-1}
\right),
\end{equation}
where $tr(A)$ stands for the trace of matrix A, and \beq \label{covJH} J(\hat\psi
)= Var[\nabla l_p(\psi;X)]|_{\psi =\hat \psi}, ~~  H(\hat\psi)= E[-\nabla^2
l_p(\psi;X)]|_{\psi =\hat \psi}, \eeq
 where $\nabla$ means the derivative with respect to $\psi$. A model with a low TIC
is a good one.

\section{Bayesian model averaging for extreme events}

\subsection{Bayesian model averaging using GEV distribution}

Over the past few decades, ensemble forecasts based on global climate models have become an important part of climate forecasting
because of their ability to reduce prediction uncertainty.
 Among the various ensemble methods, BMA combines the forecast distributions of different models and builds
 a weighted predictive distribution from them.
 Many empirical studies including Raftery et al. (2005), Sloughter et al. (2007), Wang et al. (2012), Mok et al. (2018), and Huo et al. (2018)  have shown that various
 BMA approaches outperform other competitors in prediction performance.
 To calculate the weight of each forecast model, Raftery et al. (2005) used
 the expectation maximization algorithm and estimated the weights based on the performance of each model during a training period.

In standard BMA, the conditional distribution of each individual model is
assumed to follow a normal distribution, which is valid for temperature and
sea-level pressure, for example. For other variables such as maximum
precipitation or extreme wind speed, the GEVD might be a better
alternative for representing the distribution of the model's output. In their climate change
study of extreme rainfall, Zhu et al. (2013)
 used the GEVD in a BMA framework,
where the weight of each forecast model was calculated by comparing the
reanalysis data with the historical data from a simulation model. Moreover they
applied the bootstrap technique instead of employing the expectation maximization algorithm to compute
the weights of the models and variance of the BMA predictions of rainfall
intensity.
 Now, we describe the details of their method, which is referred
to as the GEV-BMA in this paper.

Given the relatively short record of 30 years from the climate model and
reanalysis data in this study, the parameter estimation of the GEVD may be
imprecise and problematic for predicting events beyond the range of the observed
data (e.g., return level for 100 years). Thus, Zhu et al. (2013) proposed a bootstrap-based approach to establish the uncertainty of the prediction. Based on the ability of each climate model's
 historical runs in matching the reanalysis data for each location, the weights for all RCMs were calculated.
 For each RCM and each bootstrap realization, the GEV parameters were generated and rainfall intensity (i.e., return level) in return period T estimated.
 The variance in rainfall intensity was calculated for all bootstrap realizations
 (from i=1 to B). They then used the Gaussian likelihood function as follows:
\begin{equation} \label{likeMk}
L(M_k, T)=\frac{1}{\sqrt{2\pi}\sigma_I(T)}exp\left[-\frac{\frac{1}{B}\sum_{i=1}^{B}[I_i(T)-I_i^k(T)]^2}{2\sigma^2_I(T)}\right],
\end{equation}
where $I_i(T)$ is the intensity of the i-th bootstrap from the reanalysis data, $I_i^k(T)$ is
the intensity of the i-th bootstrap from the historical data of model $M_k$, and
\begin{equation}\label{sigmaI}
\sigma_I^2(T)=\frac{1}{B}\sum_{i=1}^{B}[I_i(T)-\bar{I}(T)]^2
\end{equation}
is the variance in intensity based on the reanalysis data, where $\bar{I}(T) =
\frac{1}{B}\sum_{i=1}^{B} I_i(T)$. Note that the data from all models were
downscaled to have the same grids as the reanalysis data in advance of fitting
the GEVD at each site.

If $I=I(T)$ is the rainfall intensity of return period $T$ predicted by a set
of K alternative models, then its distribution conditioned on the reanalysis data
$\textbf{R}$ is (Hoeting et al., 1999)
\begin{equation}
p(I|\textbf{R})=\sum_{k=1}^{K}p(I|M_k, \textbf{R})p(M_k|\textbf{R}),
\end{equation}
where $p(M_k|\textbf{R})$ is the predictive probability of $I$ for model $M_k$
and $p(M_k|R)$ is the posterior probability of $M_k$. One way of estimating the
posterior model probability is to use Bayes' theorem:
\begin{equation}\label{wgt}
p(M_k|\textbf{R})=\frac{p(\textbf{R}|M_k)p(M_k)}{\sum_{l=1}^{K}p(\textbf{R}|M_l)p(M_l)},
\end{equation}
where $p(\textbf{R}|M_k)$ is the likelihood of model $M_k$ and $p(M_k)$ is the
prior probability of $M_k$. The above quantity (\ref{wgt}) is used as the weights of
the BMA as in the below equation (\ref{bmapred}). Here, $p(M_k)$ may be
assigned based on prior information from expert elicitation, professional
assessments of the incorporated models, and so on. If such information is unavailable,
the prior probabilities could be set to be equal for all the models. In the numerical
study of this article, equal priors are used for all 17 models. The likelihood of
model $M_k$ is approximated by using the generalized likelihood uncertainty
estimation method (Beven, 2006; Rojas et al., 2008):
\begin{equation}
p(\textbf{R}|M_k)=L(M_k).
\end{equation}

The mean and variance of the BMA prediction of rainfall intensity $I$ over $K$
models are then given by
\begin{equation} \label{bmapred}
E(I|\textbf{R})=\sum_{k=1}^{K}E[I|\textbf{R}, M_k]w_k,
\end{equation}
\begin{equation} \label{postvar}
Var(I|\textbf{R})=\sum_{k=1}^{K}[E(I|\textbf{R},M_k)-E(I|\textbf{R})]^2w_k+\sum_{k=1}^{K}Var[I|\textbf{R}, M_k]w_k,
\end{equation}
where $w_k=p(M_k|\textbf{R})$ is the weight for each model $k$.
For the historic data, these quantities are estimated by the calculation from the $B$
bootstrap samples used in (\ref{likeMk}): \beq
\label{est-pred}
 \hat E[I|\textbf{R}, M_k] =
\bar{I}(T, M_k ) = \frac{1}{B}\sum_{i=1}^{B} I_i^k(T)
\eeq
and
\beq \label{est-var}
{\widehat Var}[I|\textbf{R}, M_k] = ~ \sigma_I^2(T, M_k )
=\frac{1}{B}\sum_{i=1}^{B}[I_i^k(T)-\bar{I}(T, M_k)]^2. \eeq
For the future simulation data, the above equations (\ref{est-pred}) and (\ref{est-var}) are not used, but the intensity estimates obtained by fitting GEVDs to each model's future data are used. That is, $\hat{E}(I|R,M_k) = \hat{I}^k_f(T)$, where the subscript $f$ stands for the future data. For estimation of $Var(I|R,M_k)$, we may need to consider other quantities such as asymptotic variance.



\subsection{Bayesian model averaging using MSP} \label{sec_boot}

In Zhu et al. (2013), the dataset from a simulation model in each grid point was
treated independently even though nearby observations were
aggregated. That is, the GEVD was fitted at-site and the return levels
in period T were estimated for each grid. The weighted average of these return
levels over a suite of simulation models is the GEV-BMA prediction. As
mentioned in subsection \ref{msp}, the precision of these estimated
return levels can be improved by applying the MSP instead of using
point-wise GEV models. Now, we propose an integration of BMA and the MSP with the
expectation of an improved ensemble prediction compared with the GEV-BMA. We refer to this new integrated approach as the MSP-BMA.

In the MSP-BMA, the MSPs are fitted to the reanalysis data and the historical data of
model $M_k$, respectively, over the spatial domain. Then, the return level,
 $I(T)$, at each site $x$ is obtained by using the following equation:
 \beq \label{msp-zp}
I(T)_x = \mu(x) - {\sigma(x) \over \xi (x) } [ 1- \{ - log (1-p) \} ^{-\xi (x)} ],
\end{equation}
where $p =1/T$. This is a spatial extension of equation (\ref{z_p}). Here, the GEV
parameters $\mu(x), \;\sigma(x),\; \xi(x)$ obtained from the MSP fitting are
functions of the spatial variables given in (\ref{para-model}) in the next section. To
compute (\ref{likeMk}), bootstrap samples are used as in the GEV-BMA. These
are obtained by random sampling with replacement over the spatial domain, from
the reanalysis data and the historical data of model $M_k$. It was sampled
`year-wise' over spatial stations. If the year 1980 is selected, for example, then
the observation vector of all stations on the year is used as a bootstrap
realization. Then, the BMA weights are calculated by using the same formula as
in (\ref{wgt}). Moreover, the BMA prediction and uncertainty estimates over
multi-models for the future data are calculated by the same equations as in
(\ref{bmapred}) and (\ref{postvar}), respectively.

The main advantage of the MSP-BMA is that the precision of the BMA prediction
is improved, as shown by the illustrative numerical study in the next section.
As an advantage of MSP-BMA as a by-product,  we have
experienced a technical convenience of handling data in applying MSP-BMA to
multiple climate models compared to applying GEV-BMA. It is because that
each climate model has different resolution and so has different locations and
sizes of grid cells. Moreover, those are also different from those of reanalysis
data. Thus, to compare each model to reanalysis data and to combine
multi-models, we needed a pre-processing to make the center points of grid cells
to be same (i.e., regridding) across multi-models and reanalysis data in applying
GEV-BMA, while we do not need it in applying MSP-BMA.

\subsection{Comparison methods}

To compare the performances of the GEV-BMA and the MSP-BMA, we
considered the following two criteria. The first one is the quantile per quantile (qq) plot for the group-wise maxima to check the goodness-of-fit (Davison and
Gholamrezaee, 2012). Here, the empirical group-wise maxima for each year are
selected from the given observations across the sites, while the theoretical
group-wise maxima are obtained from the average of the $B$ randomly generated
MSPs for each year. This is done for all the years (30 years in
this study) to construct a maxima series. Then, the quantiles from each maxima
series are plotted. This plot with a confidence band is drawn for each method and
compared.

The second criterion is the variance in return levels of each method for each
site. This is calculated by using equation (\ref{est-var}) for both methods. Lower
variance values represent less uncertainty in the estimation. This comparison is conducted for each
site (10 locations in this study). The relative improvement is calculated by
\begin{equation} \label{relImp}
{{Var (\text{GEV-BMA}) - Var (\text{MSP-BMA})} \over {Var (\text{GEV-BMA})}} \times 100.
\end{equation}

\section{Numerical study} \label{sec:NS}
\subsection{Datasets}

We analyze annual maximum daily precipitation (AMP1) as simulated by
the CMIP5 models in the historical experiments (1971--2000) and the experiments
for the 21st century (2021--2050). We considered the models only
employing one radiative forcing scenario, called Representative Concentration Pathways (RCP) 4.5.  Table \ref{tab-CMIP5} lists the model names, institutes,
and resolution of the 17 models considered in this study.

\begin{figure} [tbh]
\begin{center}
\includegraphics[width=8cm, height=8.5cm]{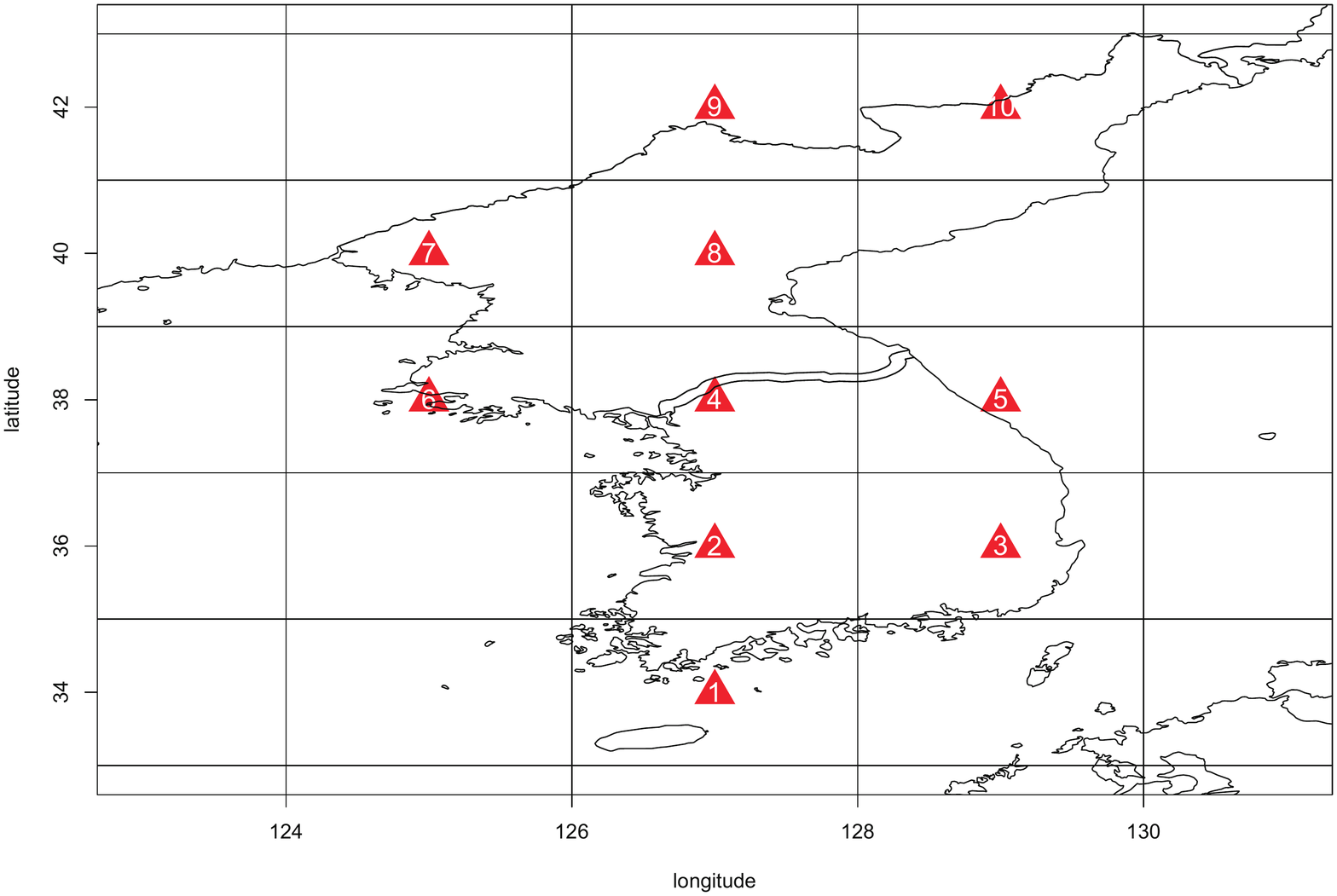}
\caption{Map of the Korean peninsula with grids of $2^o \times 2^o$ and {central} points in each box.
\label{map_korea}}
\end{center}
\end{figure}

For the reanalysis data, we use APHRODITE (Asian Precipitation Highly-Resolved
Observational Data Integration Towards Evaluation, v1101) daily gridded precipitation
data. This is the long-term (1951 onward) continental-scale daily product that
contains a dense network of daily rain-gauge data for Asia including the
Himalayas, South and Southeast Asia, and mountainous areas in the Middle
East (Yatagai et al., 2009). The grid resolution of APHRODITE is $0.5^o \times
0.5^o$ for East Asia including Korea. We treat this reanalysis data as the ``ground
truth.''

In a statistical analysis such as the GEV-BMA, we need to match the grid point
of the reanalysis data to that of the CMIP5 models. However, the resolution of
the CMIP5 models varies, as seen in Table \ref{tab-CMIP5}, and thus we need a
standard grid to efficiently use these datasets. For such a standard grid, we set
the grids as $2^o \times 2^o$, as seen in Figure \ref{map_korea}. For each box
in the figure, there are 16 observations of AMP1 from APHRODITE. We treat the
average of these 16 observations as the representative AMP1 of the grid box. For
our study, 10 boxes are chosen over the Korean peninsula. The red triangle in
Figure \ref{map_korea} stands for the center site of the box. For each site, a time
series of 30 years (from 1971 to 2000) AMP1 is used. For the historical and
future data from each simulation model, {the Kriging method is
applied to obtain the observation (AMP1) at the center site. The ``DiceKriging''
package in R software (Roustant et al., 2012) was used with Matern 5/2
covariance structure.}
 Now, we have three kinds of datasets for
each site: reanalysis and historical data from 1971 to 2000 (period 0) and future
simulation data from 2021 to 2050 (period 1).

\begin{table}[htbp]
  \centering
  \caption{The list of the 17 CMIP5 climate models analyzed in the present study and their horizontal and vertical resolutions.
  Model resolution is characterized by the size of a horizontal grid on which output is available from the model's atmospheric
  component and by the number of vertical levels.
  Spectral models are also characterized by their spectral truncations in brackets.}
  {
      \begin{tabular}{lll}
      \hline
\hline
 {Model Name} & \multicolumn{1}{l}{Institution} & \multicolumn{1}{l}{Resolution} \\
&&(Lon$\times$Lat Level\#)\\\hline
    \multirow{1}[0]{*}{BCC-CSM1-1} &Beijing Climate Center, China Meteorological Ad-& \multirow{1}[0]{*}{192$\times$145L38} \\
&ministration, China \\

  \multirow{1}[0]{*}{CanESM2} &Canadian Centre for Climate Modelling and Anal-& \multirow{1}[0]{*}{128$\times$64L35(T63)} \\
&ysis\\

    \multirow{1}[0]{*}{CCSM4} &National Center for Atmospheric Research(NCAR),& \multirow{1}[0]{*}{288$\times$192L26} \\
&USA\\

   \multirow{1}[0]{*}{CNRM-CM5} &Centre National de Recherches Meteorologiques,& \multirow{1}[0]{*}{256$\times$128L31(T127)} \\
&Meteo-France\\

   \multirow{1}[0]{*}{GFDL-ESM2G} &Geophysical Fluid Dynamics Laboratory, USA& \multirow{1}[0]{*}{144$\times$90L24} \\

   \multirow{1}[0]{*}{GFDL-ESM2M} &Geophysical Fluid Dynamics Laboratory, USA& \multirow{1}[0]{*}{144$\times$90L24} \\

    \multirow{1}[0]{*}{HadGEM2-AO} & National Institute of Meteorological Research/ & \multirow{1}[0]{*}{192$\times$145L40} \\
 & Korea Meteorological Administration, Korea  \\

    \multirow{1}[0]{*}{HadGEM2-CC} &UK Met Office Hadley Centre& \multirow{1}[0]{*}{192$\times$145L40} \\

    \multirow{1}[0]{*}{HadGEM2-ES} &UK Met Office Hadley Centre& \multirow{1}[0]{*}{192$\times$145L40} \\

    \multirow{1}[0]{*}{INMCM4} &Institute for Numerical Mathematics, Russia & \multirow{1}[0]{*}{180$\times$120L21} \\

    \multirow{1}[0]{*}{IPSL-CM5A-LR} &Institute for Pierre-Simon Laplace, France & \multirow{1}[0]{*}{96$\times$96L39} \\

    \multirow{1}[0]{*}{MIROC5} &Model for Interdisciplinary Research on Climate,& \multirow{1}[0]{*}{256$\times$128L40(T85)} \\
&Japan\\

   \multirow{1}[0]{*}{MIROC-ESM} &Model for Interdisciplinary Research on Climate,& \multirow{1}[0]{*}{128$\times$64L80(T42)} \\
&Japan\\

    \multirow{1}[0]{*}{MIROC-ESM-} &Model for Interdisciplinary Research on Climate,& \multirow{1}[0]{*}{128$\times$64L80(T42)} \\
CHEM & Japan\\

    \multirow{1}[0]{*}{MPI-ESM-LR} &Max Planck Institute for Meteorology, Germany& \multirow{1}[0]{*}{192$\times$96L47(T63)} \\

    \multirow{1}[0]{*}{MRI-CGCM3} &Meteorological Research Institute, Japan & \multirow{1}[0]{*}{320$\times$160L48(T159)} \\

    \multirow{1}[0]{*}{NorESM1-M} &Norwegian Climate Centre, Norway & \multirow{1}[0]{*}{144$\times$96L26} \\
\hline
    \end{tabular}%
    }
  \label{tab-CMIP5}%
\end{table}%

\begin{figure} [tbh]
	\begin{center}
		\includegraphics[width=13cm, height=12cm]{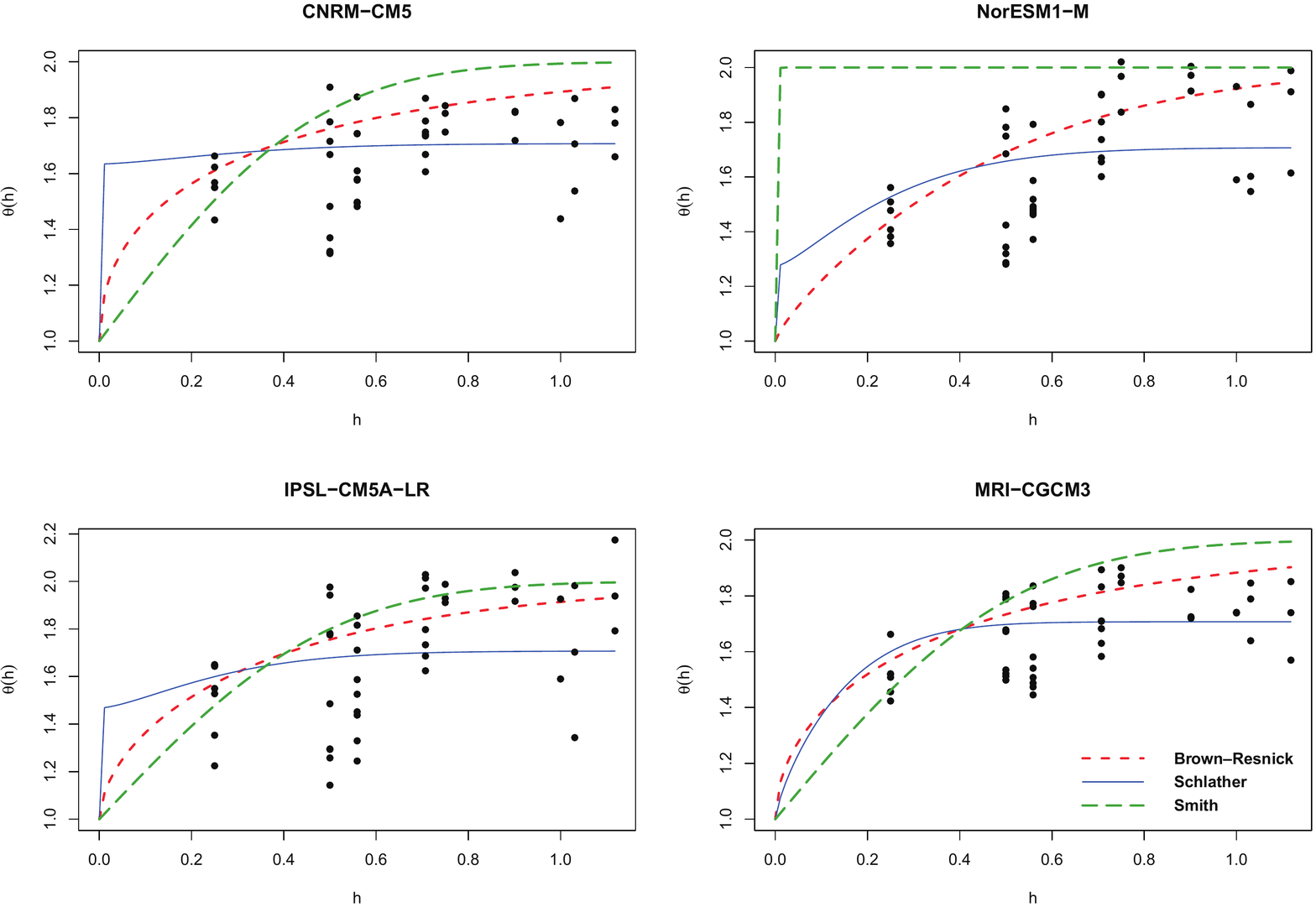}
		\caption{The extremal coefficient plots by applying three (Smith, Schlether, and Brown--Resnick) characterizations for four of the sample
			simulation models: CNRM-CM5, NorESM1-M, IPSL-CM5A-LR, and
			MRI-CGCM3.  \label{Brown-Resnick}}
	\end{center}
\end{figure}

\subsection{MSP modeling}
For the data analysis, we employ a linear transformation to make the station's
geographical information be in [0, 1]:
 \beq
   x_{new} = [{x - min(x)}]/[{max(x)-min(x)}].
\eeq
 This makes the calculation of the inverse of the
Hessian matrix $H (\psi )$ more stable. According to the data, the minimum and
maximum values of the latitudes, longitudes, and altitudes are (34.23$^{\circ}N$,
38.15$^{\circ}N$) degrees, (126.22$^{\circ}E$, 129.24$^{\circ}E$) degrees, and
(1.3m, 263.1m), respectively. To compute the MSP, we use the R
package ``SpatialExtremes'' {developed} by Ribatet et al. (2013).

From the reanalysis data, we build regression-based forms to the location and
scale parameters by including latitude and longitude, where $x$ denotes the
station. We set $\xi(x)$ as a constant $\xi$, which is typical in MSP
modeling (Buishand et al., 2008; Westra and Sisson, 2011; Lee et al., 2013), because we are studying with a small number of grid cells. The
selected model at the minimum of the TIC is as follows:
\begin{eqnarray}\label{para-model}
 \mu(x) &=& \mu_0 + \mu_{t} \; lat(x) + \mu_{g}\; long(x) +
 \mu_{t2} lat(x)^2 + \mu_{tg} \; lat(x) \times long(x), \nonumber \\
 && \sigma(x) =\sigma_0 +  \sigma_{g}\; long(x) + \sigma_{tg} \; lat(x) \times long(x),
\end{eqnarray}
where $lat(x)$ and $long(x)$ are the latitude and longitude of the site $x$ and
$\mu_0, \mu_t,  \mu_g$, $ \mu_{t2}, \mu_{tg}$, $\sigma_{0}, \sigma_{g},
\sigma_{tg}$ are the regression coefficients corresponding to the variables. This
selected regression form is fixed for the MSP for the historical and future data in this
study because the model selection based on the TIC for the 17 CMIP5 models took
huge computing time. However, the regression coefficients are estimated for every
dataset by maximizing the pair-wise log likelihood function given in (\ref{CL}).
Here it is notable, as commented by a reviewer, that the polynomial form for GEV parameters is just a convenience assumption but may not be very flexible.
Treating GEV parameters as Gaussian processes in spatial Bayesian hierarchical model may provide a more flexible alternative.

\begin{figure} [tbh]
\begin{center}
\includegraphics[width=12cm, height=7.5cm]{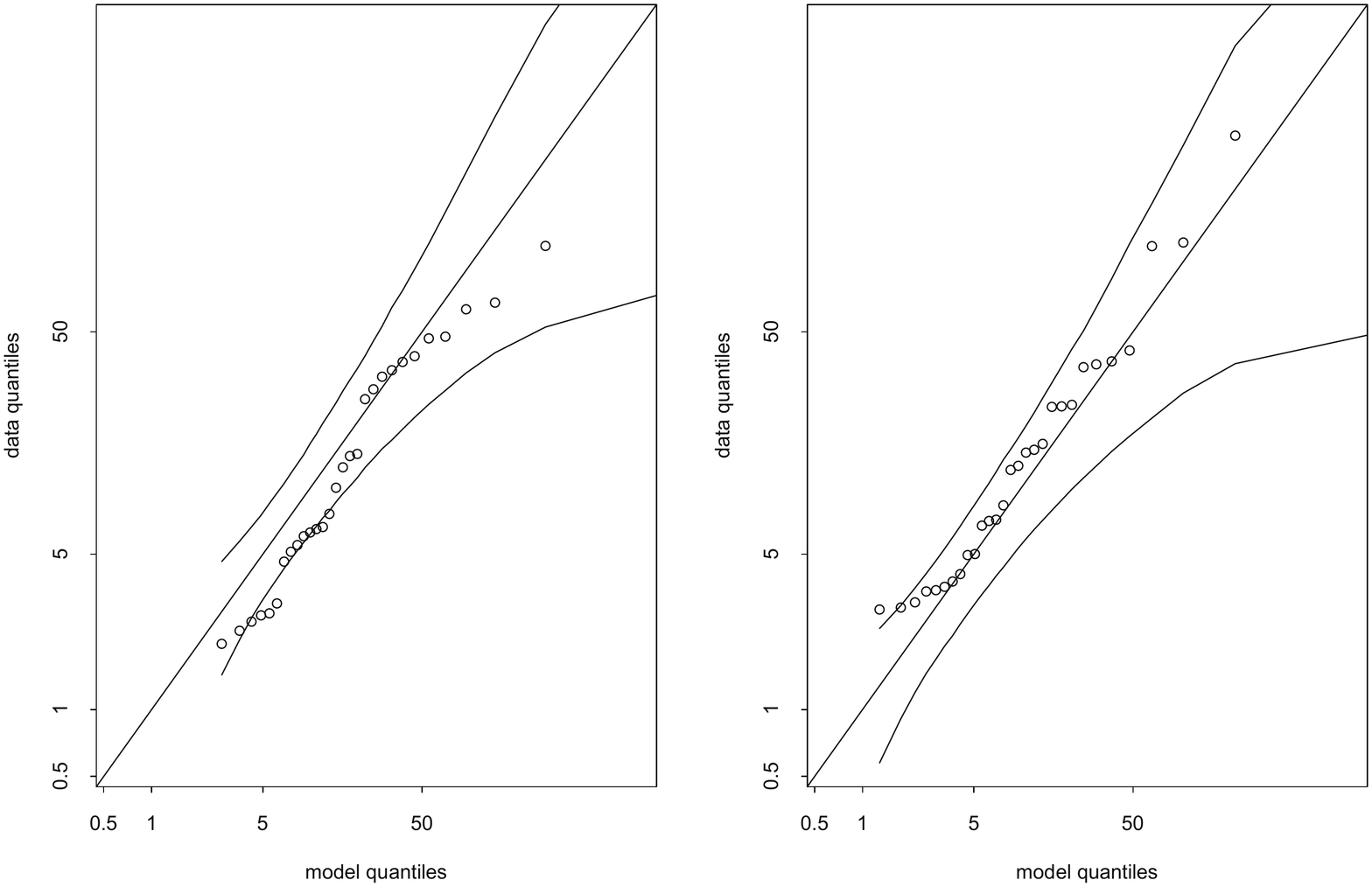}
\caption{Group-wise maxima qq plots for the reanalysis data. The left panel is from the
GEVD fitting and the right one is from the MSP fitting.
\label{fig:gmqq} }
\end{center}
\end{figure}

To select the characterization among the three approaches (Smith,
Schlather, and Brown--Resnick), we check the extremal coefficient plots for
each simulation model. Figure \ref{Brown-Resnick} shows typical plots drawn
for some of the simulation models: CNRM-CM5, NorESM1-M, IPSL-CM5A-LR, and
MRI-CGCM3. From these plots, we choose the Brown--Resnick
characterization in this study.

Figure \ref{fig:gmqq} shows the group-wise maxima qq
plots. The left panel is from the GEV fitting and the right one is from the MSP fitting to the
reanalysis data. The right plot provides a better qq than the left one in the sense
that the points in the right plot are closer to the straight line and more
located inside the confidence band. This means that the MSP is more acceptable
than the GEVD for this dataset.

\subsection{BMA result for comparison}

\begin{figure} [tbh]
\begin{center}
\includegraphics[width=8cm, height=7cm]{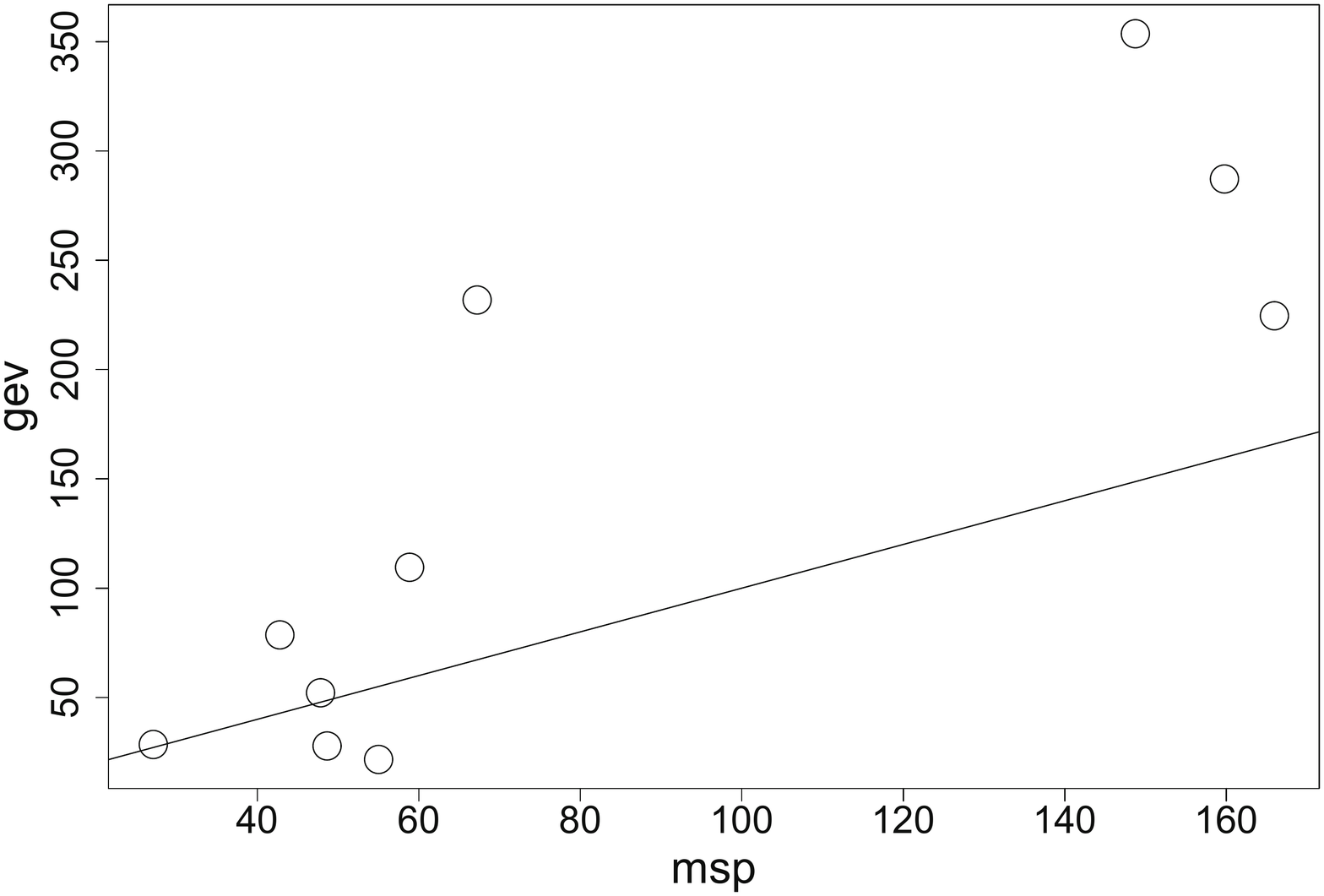}
\caption{Scatterplot of the variance estimates (unit:$mm^2$) of the 20-year return levels calculated by using both BMA methods.
The Y-axes and X-axes stand
for the variances estimated by using the GEV-BMA and the MSP-BMA, respectively.
\label{vardiff}}
\end{center}
\end{figure}

For each site, the variance estimates of the 20-year return levels are calculated from the
GEV-BMA and the MSP-BMA by using the reanalysis and historical data. Figure \ref{vardiff}
compares these two kinds of variances (unit: $mm^2$). The Y-axes and X-axes stand
for the variances estimated by using the GEV-BMA and the MSP-BMA, respectively.
The GEV-BMA has greater variance than the MSP-BMA for eight of the 10 sites. This may mean that the MSP-BMA provides more precise predictions with less uncertainty
than the GEV-BMA. Table \ref{tab:RelImp} shows the estimated variances of
the 20-year return levels obtained by using both BMA methods. The relative improvement is calculated by the formula (\ref{relImp}) in which the MSP-BMA achieves a 42\% relative variance reduction on average over the GEV-BMA.

\begin{table}[ht]
\caption{Variance estimates (unit:$mm^2$) for the 20-year return levels calculated for the reanalysis data for each site by using both BMA methods. The relative improvement is calculated by the formula (\ref{relImp}).}
\vskip .3 cm
\centering
\begin{tabular}{rrrc}
\hline
Site& Var(GEV-BMA) & Var(MSP-BMA) & Relative Improvement(\%) \\
\hline
1 & 353.6 & 148.7 & 57.9  \\
2 & 109.5 & 58.9  & 46.3  \\
3 & 224.6 & 166.0 & 26.1  \\
4 & 231.8 & 67.2  & 71.0  \\
5 & 287.2 & 159.8 & 44.4  \\
6 & 78.6  & 42.8  & 45.6  \\
7 & 27.8  & 48.6  & -74.9 \\
8 & 21.6  & 55.0  & -154.3 \\
9 & 28.5  & 27.1  & 4.8   \\
10& 52.1  & 47.8  & 8.3   \\
\hline
 Average& 141.5 & 82.2  & 41.9 \\
\hline
\end{tabular}  \label{tab:RelImp}
\end{table}

It may not be surprise that the variances obtained from MSP is smaller than point-wise GEV,
but it could also due to the spatial modeling bias and the variance estimation may be too small to be true. Thus we used a bootstrap method to estimate the bias more reasonably.
 Bootstrap samples are obtained by random sampling with replacement over the spatial domain, from
the reanalysis data and the historical data of model $M_k$. It was sampled
`year-wise' over spatial stations, as we did in the previous section \ref{sec_boot}.
The bias is calculated by the difference between the true 20-year return level  and estimated levels obtained by a bootstrap sample. This calculation is repeated $B$ times, and the average of $B$ differences is considered to be the bias of the model. Here, for the true 20-year return level, we used the estimate of 20-year return level obtained by applying GEV-BMA model for the original data.  If we set
the true 20-year return level differently from this study, the bias estimates for both models may be changed. The table \ref{tab:bias_est} shows the bias estimates
calculated by using GEV-BMA and MSP-BMA methods.  
The table shows that the MSP-BMA has bias inflation over GEV-BMA at eight sites.
The bias of MSP-BMA may be introduced by assuming a regression model for each GEV parameter.
By considering both results from Tables \ref{tab:RelImp} and \ref{tab:bias_est}, we know that it is an example of bias-variance trade-off problem, 
as commented by a reviewer. We leave more study on this matter to future work.

\begin{table}[ht]
\centering
\caption{Bias for the estimate of 20-year return levels calculated by using GEV-BMA and MSP-BMA methods.}
\vskip .3cm
\begin{tabular}{cccccccc}
  \hline
Site & Bias(GEV-BMA)  & Bias(MSP-BMA) \\
  \hline
1 & 30.5 &  33.1  \\
2 & 13.9 &  17.6  \\
3 & 15.3 &  13.4  \\
4 &-14.4 &  -15.7 \\
5 & 25.2 &  27.7  \\
6 & 18.4 &  28.3  \\
7 & -7.5 & -10.8  \\
8 &-10.9 &  -20.1 \\
9 & 1.1  &  -8.2  \\
10&-11.2 &  -10.7 \\ \hline
\end{tabular} \label{tab:bias_est}
\end{table}

\section{Summary and discussion}

In this study, a MSP-embedded BMA is proposed to project extreme climates based on multi-model ensemble simulations. The proposed method, referred to as the MSP-BMA, is a spatial
extension of BMA embedded with a point-wise GEVD. The superiority of the proposed method over the GEV-embedded BMA is demonstrated by using extreme rainfall intensity data on the
Korean peninsula from CMIP5 multi-models. The APHRODITE reanalysis data and 17 CMIP5 models are
examined for 10 grid boxes over the Korean peninsula.

A conclusion of this study is that
 the MSP-BMA outperforms the GEV-embedded BMA by generating lower prediction variance estimates
 for extreme rainfall intensity data on the Korean peninsula.
 The MSP-BMA achieves a 42\% relative variance reduction over the GEV-embedded BMA on average. However, the regression-based forms in MSP may introduce some bias, which needs more study on this bias-variance trade-off problem.

 A by-product technical advantage of the MSP-BMA
is that tedious `regridding' is not required before and after the analysis while it should be done for the GEV-embedded BMA.

As commented by a reviewer, the constant $\xi$ parameter assumption in MSP effectively reduces
the estimation uncertainty while there is no such a constraint from point-wise GEV method. For a fair comparison,
one can also modify the point-wise GEV method to a constant shape parameter GEV model (Wang et al. 2016).
We leave this matter to future work.

\renewcommand{\baselinestretch}{1.0}
\section*{Acknowledgment}
\small
The authors would like to thank two reviewers and Associate Editor for their valuable comments and constructive
suggestions. This work was supported by the
National Research Foundation of Korea (NRF) grant funded by the Korean
government (No.2016R1A2B4014518), and also funded by the Korea Meteorological Administration Research and Development Program under Grant KMI2018-03414. Lee's work was supported by the
Basic Science Research Program through the NRF funded by the Ministry of Education (2017R1A6A3A11032852).


\begin{thebibliography}{100}
	\small

\bibitem[Arisido et al. (2017)]{Arisido}
Arisido, M.W., Gaetan, C., Zanchettin, D. et al. A Bayesian hierarchical approach for spatial analysis of climate model bias in multi-model ensembles.
Stoch Environ Res Risk Assess (2017) 31(10):2645--2657. https://doi.org/10.1007/s00477-017-1383-2

\bibitem[Beven (2006)]{beven}
Beven K (2006) A manifesto for the equifinality thesis. Jour of Hydrology
320:18--36.

\bibitem[{{Buishand et al.}(2008)}]{buishand}
Buishand TA, de Haan L, Zhou C (2008) On spatial extreme: With
application to a rainfall problem. Annals of Applied Statistics 2(2):624--642.

\bibitem[Casey (1995)]{Casey}
Casey T (1995) Optimal linear combination of seasonal forecasts. Austral. Meteor. Magazine 44:219--224.

\bibitem[Coelho (2004)]{Coelho}
Coelho CA, Pezzulli SS, Balmaseda M et al. (2004) Forecast calibration and combination: A simple Bayesian approach for ENSO. Jour Climate 17:1504--1516.

\bibitem[{{Coles}(2001)}]{coles}
Coles S (2001) An Introduction to Statistical Modelling of
Extreme Values. Springer, New York, pp 224.

\bibitem[Cooley (2012)]{Cooley}
Cooley D, Cisewski J, Erhardt RJ, Jeon S et al. (2012) A survey of spatial extremes: Measuring spatial dependence and modeling spatial effects. Revstat 10(1):135--165.

\bibitem[{{Cressie and Wikle}(2011)}]{cressie11}
Cressie N, Wikle C (2011) Statistics for spatio-temporal data. Wiley, Hoboken.

\bibitem[{{Davison and Gholamrezaee}(2012)}] {DavisonGhol}
Davison AC, Gholamrezaee MM (2012) Geostatistics of extremes. Proc Royal Society A  468:581--608. doi:10.1098/rspa.2011.0412.

\bibitem[{{Davison et al.}(2012)}] {DavisonPR}
Davison AC, Padoan SA, Ribatet M (2012) Statistical modelling of spatial
extremes (with discussions). Statistical Science 27(2):161--186.


\bibitem[{{de Haan}(1984)}] {deHaan84}
 de Haan, L (1984) A spectral representation for max-stable
 processes. Annals of Probability 12(4):1194--1204.


\bibitem[{{de Hann and Pereira}(2006)}]{deHaanP}
 de Haan L, Pereira TT (2006) Spatial extremes: models for the stationary
 case. Annals of Statistics 34:146--168.

 \bibitem[Diggle (2007)]{Diggle}
 Diggle PJ, Ribeiro PJ Jr. (2007) Model-based Geostatistics. Springer, New York.


 \bibitem[Fawcett (2016)]{Fawcett}
 Fawcett L, Walshaw D (2016) Sea-surge and wind speed extremes: optimal estimation strategies for planners and engineers. Stoch Environ Res Risk Assess 30(2):463–-480.

\bibitem[{{Gaume et al.}(2013)}]{Gaume}
Gaume J, Eckert N, Chambon G, Naaim M, Bel L (2013) Mapping
extreme snowfalls in the French Alps using max-stable processes. Water
Resour Res 49. doi:10.1002/wrcr.20083


\bibitem[Hoeting et al.(1999)]{hoeting99}
Hoeting JA, Madigan D, Raftery AE, Volinsky CT (1999) Bayesian model
averaging: a tutorial. Statistical Sciences 14(4):382--417.


\bibitem[Hosking (1990)]{hosking90}
Hosking JRM (1990) L-moments: analysis and estimation of distributions using
linear combinations of order statistics. Journal of the Royal Statistical
Society, Series B 52:105--124

\bibitem[Hosking et al. (1985)]{hosking85}
Hosking JRM, Wallis JR, Wood EF (1985). Estimation of the generalized extreme-value distribution
by the method of probability weighted moments. Technometrics 27(3):251–261.



\bibitem [Huo et al.(2018)]{Huo}
Huo W, Li Z, Wang J et al. (2018) Multiple hydrological models comparison and an improved Bayesian model averaging approach for ensemble prediction over semi-humid regions. Stoch Environ Res Risk Assess. https://doi.org/10.1007/s00477-018-1600-7

\bibitem[{{Kabluchko \it{et al.}}(2009)}]{Kabluchko}
Kabluchko Z, Schlather M, de Haan L (2009) Stationary
max-stable fields associated to negative definite functions. Ann Probab 37:2042--2065

 \bibitem[Kharin et al.(2013)]{Kharin13}
 Kharin VV, Zwiers FW, Zhang X, Wehner M. (2013). Changes in temperature and precipitation extremes in the CMIP5 ensemble.
 Climatic Change 119:345-357.



\bibitem[Leadbetter (1983)]{Leadbetter}
Leadbetter MR, Lindgren G, Rootzen H (1983) Extremes and Related Properties of Random Sequences and Processes. Springer Verlag, 336 pp.

\bibitem[{{Lee et al.}(2012)}] {LeePark}
Lee Y, Yoon S, Murshed MS, Kim MK, Cho CH, Baek HJ, Park JS (2012) Spatial modeling of the highest daily maximum temperature in Korea
via max-stable processes. Advances in Atmospheric Sciences 30(6):1608--1620


\bibitem[Madadgar and Moradkhani (2014)]{madadgar14}
Madadgar S, Moradkhani H (2014) Improved Bayesian multimodeling: Integration
of copulas and Bayesian model averaging. Water Resource Res 50(12):9586--9603

\bibitem[Mok (2018)]{Mok}
Mok KM, Yuen KV, Hoi KI et al. (2018) Predicting ground-level ozone concentrations by adaptive Bayesian model averaging of statistical seasonal models. Stoch Environ Res Risk Assess  32(5):1283--1297. https://doi.org/10.1007/s00477-017-1473-1

\bibitem[Najafi and Moradkhani (2015)]{najafi15}
 Najafi MR, Moradkhani H (2015) Multi-model ensemble analysis of runoff extremes for climate change impact assessments. Journal of Hydrology 525:352--361

 \bibitem[Nelsen (2006)]{Nelsen}
 Nelsen RB (2006) An Introduction to Copulas, 2nd ed. Springer, New York.

\bibitem[Oesting and Stein (2017)]{Oesting}
Oesting M, Stein A (2017) Spatial modeling of drought events using max-stable
processes. Stoch Environ Res Risk Assess. Online First doi:10.1007/s00477-017-1406-z

\bibitem[{{Padoan et al.}(2010)}] {Padoan}
Padoan SA, Ribatet M, Sisson SA (2010) Likelihood-based
inference for max-stable processes. Journal of the American
Statistical Association 105:263--277.

\bibitem[Parrish et al.(2012)]{parrish12}
Parrish MA, Moradkhani H, DeChant CM (2012) Toward reduction of model uncertainty: Integration of Bayesian model averaging and data assimilation. Water Resource Res 48:W03519.

\bibitem[{{Raftery et al.}(2005)}]{Raftery}
Raftery AE, Gneiting T, Balabdaoui F, Polakowski M (2005) Using Bayesian model averaging to calibrate forecast ensembles. Monthly Weather Review 133(5):1155--1174.

\bibitem[Requena (2016)]{Requena}
Requena, A.I., Flores, I., Mediero, L. et al. (2016) Extension of observed flood series by combining a distributed hydro-meteorological model and a copula-based model.  Stoch Environ Res Risk Assess 30(5):1363--1378. https://doi.org/10.1007/s00477-015-1138-x

\bibitem[{{Ribatet et al.}(2013)}]{ribatet-guide}
Ribatet M, Singleton R, Team RC (2013) SpatialExtremes: Modelling Spatial Extremes. R package version 2.0-1. http://spatialextremes.r-forge.r-project.org/docs/SpatialExtremesGuide.pdf
[accessed 10 Jan 2017]

\bibitem[Ribatet (2013)]{ribatet13}
Ribatet M (2013) Spatial extremes: Max-stable processes at work. Journal
of French Statistical Society 154(2):156--177

\bibitem[Rojas et al.(2008)]{rojas}
Rojas R, Feyen L, Dassargues A (2008) Conceptual model uncertainty in
groundwater modeling: combining generalized likelihood uncertainty estimation
and Bayesian model averaging. Water Resour Research 44:W12418.
doi:10.1029/2008WR006908

\bibitem[{{Roustant \it{et al.}}(2012)}]{Roustant}
Roustant O, Ginsbourger D, Deville Y (2012) DiceKriging, DiceOptim: Two
R packages for the analysis of computer experiments by kriging-based
metamodeling and optimization. Jour Statist Software 51(1):1--55

\bibitem[{{Sang and Gelfand}(2010)}]{sang}
Sang, H. Y., and A. E. Gelfand, 2010: Continuous Spatial Process
Models for Spatial Extreme Values, {\it Jour. Agric. Bio. Envir.
Stat,} \textbf{15}, 49--65.

\bibitem[Sabourin et al.(2013)]{Sabourin}
Sabourin A, Naveau P, Fougeres, AL (2013) Bayesian model averaging for
multivariate extremes. Extremes 16(3):325--350

\bibitem[Salvadori (2004)]{Salvadori}
Salvadori G, De Michele C (2004)
Frequency analysis via copulas: Theoretical aspects and applications to hydrological events.
Water Resour. Res.  40(12):W12511.


\bibitem[{{Schlather}(2002)}]{schlather}
Schlather M (2002) Models for stationary max-stable random
fields. Extremes 5:33--44

\bibitem[Sloughter et al.(2007)]{sloughter}
Sloughter JM, Raftery AE, Gneiting T, Fraley C (2007) Probabilistic
quantitative precipitation forecasting using Bayesian model averaging. Monthly Weather Review 135:3209--3220


\bibitem[{{Smith}(1990)}]{smith}
Smith RL (1990) Max-stable processes and spatial extremes. unpublished manuscript.
http://www.stat.unc.edu/postscript/rs/spatex.pdf [accessed 5 Oct
2010]

\bibitem[Stephenson (2005)]{Steph}
Stephenson DB, Coelho CAS, Doblas-Reyes FJ et al. (2005) Forecast assimilation: A unified framework for the combination of multi-model weather and climate predictions. Tellus 57A:253--264.


\bibitem[{{Varin and Vidoni} (2005)}]{VarinVidoni}
 Varin C, Vidoni P (2005) A note on composite likelihood inference
 and model selection. Biometrika 92:519--528

 \bibitem[Wang1 (2016)]{Wang16}
 Wang, J., Han, Y., Stein, M. L., Kotamarthi, R., Huang W. K. (2016) Evaluation of dynamically downscaled extreme temperature using a spatially-aggregated generalized extreme value (GEV) model. Climate Dynamics 47(9):2833–2849.

 \bibitem[Wang1 (2012)]{Wang12}
 Wang QJ, Schepen A, Robertson DE (2012) Merging seasonal rainfall forecasts from multiple statistical models through Bayesian model averaging. Jour of Climate 25:5524--5537.

 \bibitem[Wang et al.(2017)]{wang17}
Wang X, Yang T, Li X et al. (2017) Spatio-temporal changes of precipitation and
temperature over the Pearl River basin based on CMIP5 multi-model ensemble.
Stoch Environ Res Risk Assess 31(5):1077--1089

\bibitem[Westra and Sisson (2011)]{westra}
Westra S, Sisson SA (2011) Detection of non-stationarity in precipitation
extremes using a max-stable process model. Jour Hydrology 406:119--128

\bibitem[Yan and Moradkhani (2016)]{Yan16}
Yan H, Moradkhani H (2016) Toward more robust extreme flood prediction by
Bayesian hierarchical and multimodeling. Natural Hazards 81(1):203--225

\bibitem[Yatagai et al.(2009)]{Yatagai09}
Yatagai A, Arakawa O, Kamiguchi K, Kawamoto H, Nodzu MI, Hamada A (2009) A 44-year daily precipitation dataset for Asia based on a dense network
of rain gauges. SOLA 5:137--140.


\bibitem [Zhu et al.(2013)]{zhu}
Zhu J, Forsee W, Schumer R, Gautam M (2013) Future projections and uncertainty assessment of extreme rainfall intensity
in the United States from an ensemble of climate models. Climatic Change 118(2):469--485. doi:10.1007/s10584-012-0639-6

\end{thebibliography}
\end{document}